\documentstyle[12pt,psfig,aaspp4]{article}
\begin{document}
\title{Fermion  stars as gravitational lenses}
\author{
Neven Bili\'c\altaffilmark{1,2},
Hrvoje Nikoli\'c\altaffilmark{1}, and Raoul D.~Viollier\altaffilmark{2}}
\altaffiltext{1}{Rudjer Bo\v{s}kovi\'{c} Institute, 10000 Zagreb, Croatia}
\altaffiltext{2}{Institute of Theoretical Physics and Astrophysics,
Department of Physics,
University of Cape Town, Rondebosch 7701, South Africa}
\begin{abstract}
We study in detail gravitational lensing
caused by a supermassive fermion star and compare
it with lensing by a black hole of the same mass.
It is argued that lensing effects, being very distinct,
may shed some light on the yet unexplained nature
of the compact dark massive object at the Galactic
center.
\end{abstract}
\section{Introduction}
\label{intro}

Compelling evidence exists
(\cite{eck96,gen96,eck97,gen97,ghe98})
that there is a
compact dark object of a mass of around
$ 2.6 \times 10^{6} M_{\odot}$,
concentrated within a radius
of $\sim$ 0.015 pc,
at the center of the Galaxy,
which is usually identified with
the enigmatic and powerful
 radio source
Sgr A$^*$.
The question
whether this is an extended object or
a supermassive black hole
is still open.
Since a compact baryonic object of that mass and size,
e.g., a cluster of low-mass stars, is almost
excluded,
it is worthwhile to explore the possibility of an
extended object made of nonbaryonic matter.

In fact, an alternative scenario for
 supermassive compact dark  objects at  galactic centers
has been  developed
in the recent past
(\cite{vio92,vio93,vio94},
\cite{tsi96,tsi98a,tsi98b,tsi99},
\cite{bil98,bil99}).
In this model,
the dark matter at the centers
of galaxies is made of nonbaryonic matter in the form
of massive neutrinos that cluster  gravitationally, forming
 supermassive neutrino stars in which the degeneracy
pressure of the neutrinos  balances their self-gravity.
Of course, the massive neutrino could be replaced by
any weakly interacting fermion of the same mass.
Such fermion stars could have been formed in the early
Universe during  a first-order gravitational
phase transition (\cite{bil97,bil98a,bil99a,bil99b}).
It has recently been shown that the dark-matter concentration
observed through stellar motion
at the Galactic center (\cite{eck97,gen96})
is consistent with a supermassive object of $2.5 \times 10^{6}$
solar masses  made of self-gravitating, degenerate heavy neutrino
matter (\cite{tsi98a}).
Moreover, it has been shown  that  an
acceptable fit to the infrared spectrum
and the radio
spectrum above 20 GHz
emitted by matter falling onto the compact dark object,
  can be reproduced
in the framework of  standard
accretion  disk theory
 (\cite{bil98,tsi99}) in
terms of a baryonic disk  immersed in the shallow potential of
 the degenerate fermion star.

In this paper, we  explore
 possible gravitational lensing effects caused by
such an object.
Gravitational lensing is a powerful tool which, if observed,
yields information on the mass distribution
of the lensing object.
\cite{war92} suggested that detection of gravitational
lensing on a 10 to 100 mas scale might be
the signature for Sgr A$^*$ being a black hole.
If the lensing object is extended and transparent to
at least part of the electromagnetic spectrum
(baryonic contamination could spoil the transparency
to a part of the spectrum),
the lensing signal would be quite distinct from
that of a  black hole.
Therefore, if a clear lensing signal is observed
around the Galactic center, it could provide a decisive
test for the black-hole or fermion-star scenarios.
For the sake of comparison, we also present some
results for stars made of weakly interacting bosons
(\cite{dab}).
These objects are in a way similar to fermion stars,
as they consist of matter that interacts
mainly gravitationally and therefore is transparent
to light, if there is no baryonic contamination.
However, boson stars have substantially different scaling
properties which make their mass and size negligible
compared with stars made of fermions.
Even a boson star with maximal mass at a distance
comparable with or larger than  stellar distances,
cannot produce a significant lensing effect
unless the boson mass is ridiculously small,
e.g., $10^{-10}$ eV.

This paper is organized as follows:
In section \ref{basics} we briefly describe
the method used and our notation.
In section \ref{comparison} we compare the
general features of lensing by a fermion star with those
by a boson star or a black hole.
In section \ref{center} we investigate
lensing effects by a hypothetical compact dark
object with properties similar to those at the Galactic center
and discuss the possibility of detecting
a lensing signal.
Our conclusions are given in section \ref{conclude}.

\section{Lensing by transparent extended objects}
\label{basics}
\begin{figure}[h]
\vbox{\vskip200pt
\includegraphics{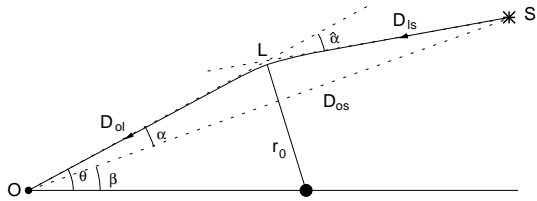}}
\caption{Geometry of  a gravitational lens system.}
\label{fig1}
\end{figure}

For a static spherically symmetric object, the metric is
\begin{equation}\label{metric}
 ds^2 =e^{\nu(r)}dt^2-e^{\mu(r)}dr^2-r^2(d\vartheta^2+\sin ^2\vartheta \,
  d\varphi^2).
\end{equation}
A ray of light, passing at a distance $r_0$ from
the center of the object, is
deflected by the angle
 (\cite{wein})
\begin{equation}\label{alkap}
 \hat{\alpha}(r_0)=2\int_{r_0}^{\infty}\frac{dr \: e^{\mu/2}}
 {\sqrt{\frac{r^4}{b^2}e^{-\nu} -r^2}} \; -\pi \; ,
\end{equation}
where $b$ is the impact parameter
\begin{equation}\label{b}
b=r_0 \exp [-\nu(r_0)/2] \; .
\end{equation}
Let $D_{ol}$, $D_{ls}$, and $D_{os}$ denote the distances from the
observer to the lens, from the lens to the source, and from the
observer to the source, respectively, as depicted in
 Figure \ref{fig1}.
The angular position of the image $\theta$ is determined by
\begin{equation}\label{theta}
\sin \theta =b/D_{ol}
\end{equation}
and (\ref{b}).
 The true angular position of the source $\beta$ is given by
the lens equation
\begin{equation}\label{beta}
 \beta =\theta -\alpha \; ,
\end{equation}
where $\alpha$ is the reduced deflection angle
\begin{equation}\label{alpha}
\alpha(r_0) ={\rm arcsin}\left( \frac{D_{ls}}{D_{os}} \sin \hat{\alpha}(r_0)
\right) \; .
\end{equation}
\vskip 10pt
The magnification of images is given by (\cite{nar})
\begin{equation}\label{mu}
\mu =\mu_t \mu_r \; ,
\end{equation}
where $\mu_t$ and $\mu_r$ are the tangential and  radial magnifications,
respectively,
\begin{equation}\label{mu2}
\mu_t =\frac{\sin \theta}{\sin \beta} \; , \;\;\;\;\;
\mu_r=\frac{d \theta}{d \beta} \; .
\end{equation}

\section{Lensing by a Schwarzschild black hole, a boson star,
and a fermion star}
\label{comparison}
We first compare gravitational lensing by
a fermion star of maximal mass with lensing by  a
black hole of the same mass and by
a boson star of maximal mass.
Since fermion and boson stars
scale in an essentially different way, we have to choose
different natural units in order to make a comparison.
The maximal masses of  boson and fermion stars are
$M_B=0.63300\; M_{\rm{Pl}}^2/m_B=
8.46 \times 10^{-11} (1\,{\rm eV}/m_B) M_{\odot}$ (\cite{lee87}) and
$M_F=0.38426\;\sqrt{2/g} M_{\rm{Pl}}^3/m_F^2=
3.43\times 10^9 \sqrt{2/g} (15\,{\rm keV}/m_F)^2 \, M_{\odot}$
(\cite{bil99b}), respectively,
where $m_B$ is the boson mass, $m_F$ the fermion mass,
and $M_{\rm Pl}=\sqrt{\hbar c/G}$ is the Planck mass.
The spin-degeneracy factor $g$ is 2 or 4 for Majorana
or Dirac fermions, respectively.
In the following numerical calculations we take $D_{ls}/D_{os}=1/2$.
Choosing  the distance
$D_{ol}=2.41732 \times 10^{5}$
 expressed in units of $\hbar /(c m_B)$,
we reproduce the results of
\cite{dab}.
In order to obtain comparable lensing caused by a maximal fermion
star, we take
$D_{ol}=2.41732 \times 10^{5}\; [a]$
 for the fermion star,
where
\begin{equation}
a=
\sqrt{\frac{2}{g}} \,
\frac{\hbar M_{\rm{Pl}}}{c \, m_F^2}
=1.3185 \times 10^{10}\,
\sqrt{\frac{2}{g}} \,
\left(
\frac{15\, {\rm keV}}{m_F}
\right)^2 \,
{\rm km},
\end{equation}
is the natural length scale
for fermion stars (\cite{bil99b}).
 We also take the same value $D_{ol}$
for lensing by a black hole.
Thus, in our comparison, the mass and the distance
expressed in physical units are equal for
the fermion star and the black hole,
and smaller by a factor
$\sqrt{2/g} M_{\rm{Pl}} m_B/m_F^2$
for the boson star.

  The  metric outside the black hole is described by the
  usual empty-space Schwarzschild solution
  (\cite{wein})
\begin{equation}\label{metric1}
e^{\nu(r)}=1-\frac{2 M_F}{r},
\end{equation}
\begin{equation}\label{metric2}
e^{\mu(r)}=\left(1-\frac{2 M_F}{r}\right)^{-1}.
\end{equation}
For the
fermion  star the metric is derived by
numerically
solving Einstein's field equations
\begin{equation}
\frac{d\nu}{dr} =
2\frac{{\cal M}+4\pi r^3 p}{r(r-2{\cal{M}})} \, ,
\label{field1}
\end{equation}
\begin{equation}
\label{field2}
\frac{d{\cal{M}}}{dr}=4\pi r^2 \rho.
\end{equation}
where the enclosed mass $\cal{M}$
 is  related to
the metric by
$e^{\mu(r)}=(1-2 {\cal{M}}(r)/r)^{-1}$,
and the pressure $p$ and the density $\rho$  are
given by the equation of state for a degenerate relativistic
Fermi gas with the metric-dependent Fermi momentum
(\cite{bil99,bil99b}).
For the boson star, the metric is derived
by solving equations (\ref{field1}) and (\ref{field2})
coupled with the Klein-Gordon equation for a
complex scalar field (\cite{lee87,dab}).

\begin{figure}[p]
\vbox{\vskip230pt
\includegraphics{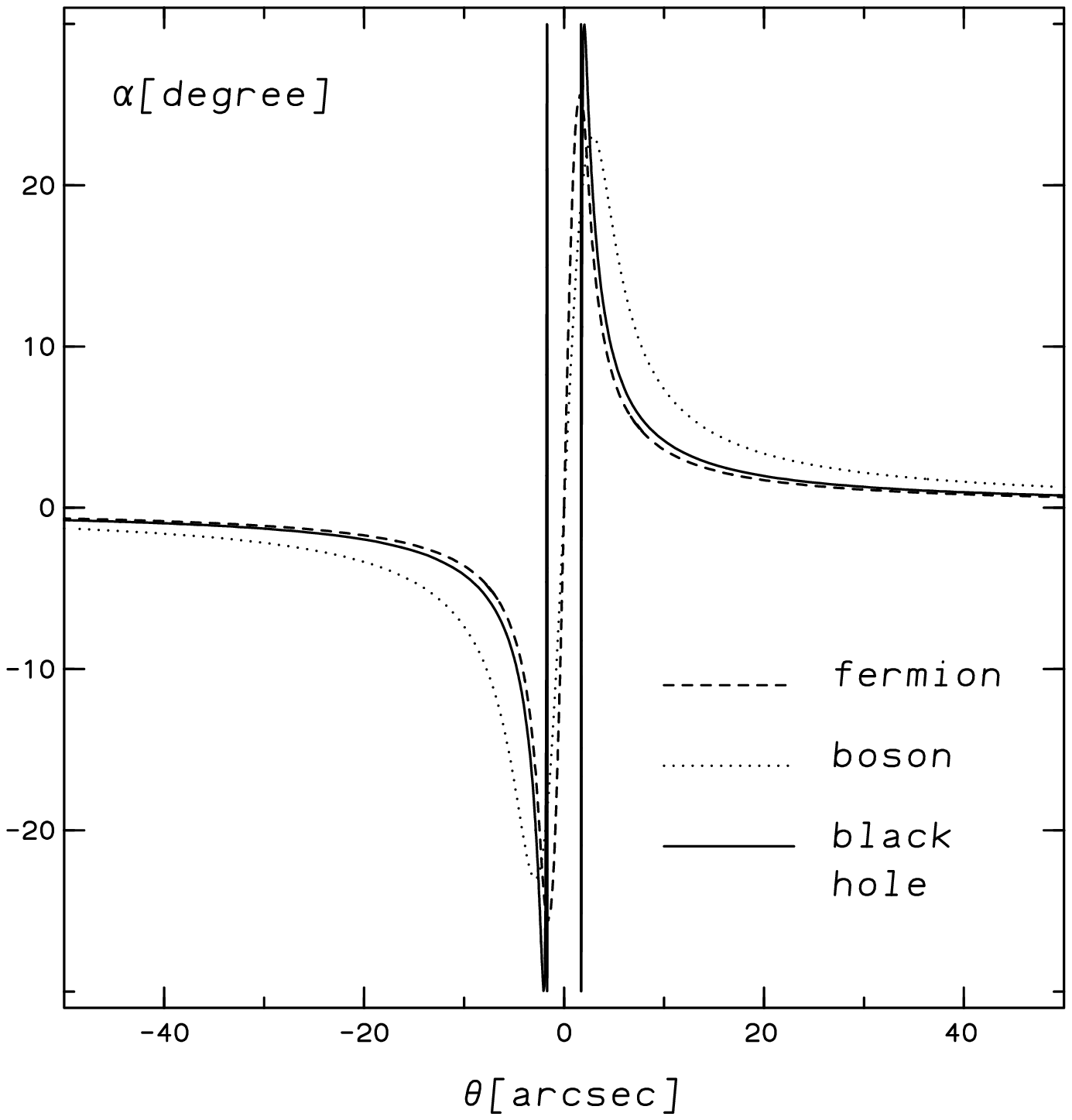}}
\caption{Reduced deflection angle versus angular position
of the image. The dotted and  dashed lines represent the maximal
boson and the fermion star, respectively,
and the solid line
represents a black hole with the same
mass as that of the fermion star.}
\label{fig2}
\end{figure}

\begin{figure}[p]
\vbox{\vskip230pt
\includegraphics{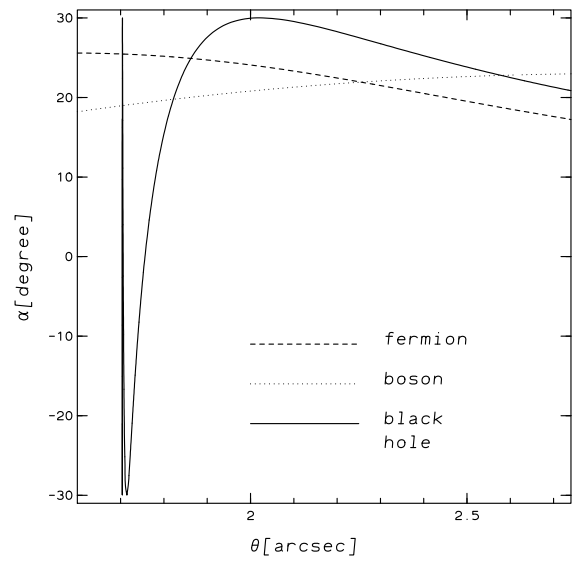}}
\caption{Same as in Figure \protect\ref{fig2},
with $\theta$ ranging from 1.6 to 2.75 arcsec.}
\label{fig3}
\end{figure}

Numerical results are presented
in Figures
\ref{fig2} and
\ref{fig3}, in which we plot the reduced deflection
angle $\alpha$ as a function of the angular
position  of the image $\theta$.
Since the masses and distances are expressed in natural
units, the plots are invariant to the choice of
$g$, $m_B$, and $m_F$.
We note qualitative similarity
between lensing by fermion and that by boson stars.
In the black-hole case
there exists a minimal angular position
$\theta_{\rm min}$ corresponding
to the impact parameter $b_{\rm min}$ such that
 the passing distance $r_0$ of the light ray
equals one and a half Schwarzschild radii $R_{\rm S}$.
 That gives (in natural units)
 $ b_{\rm min}= \sqrt{27} M_F $,
so that $\theta_{\rm min} = b_{\rm min}/D_{ol}
  = 1.7$ arcsec.
  The function $\alpha(\theta)$
 begins to oscillate near the forbidden region
 between -1.7 and 1.7 arcsec, indicating multiple images
 appearing approximately
  $1.5\, R_{\rm S}$ away
 from the center.
Each oscillation corresponds to a single  winding
 of the light ray  around the black hole.
In contrast to that,
both the boson and the fermion star curves have a smooth
transition in the central region.
The part of the curve that passes through the center
of the plot   in Figure \ref{fig2}
corresponds to a secondary image as a result of the
ray of light that goes through the object.

\section{Lensing by Sgr A$^*$}
\label{center}
In this section
 we study  possible lensing effects caused by
a supermassive dark object at the center of
our galaxy, which we further refer to as Sgr A$^*$.
In particular,   the trajectories
of the images are examined assuming that a lensed star
behind Sgr A$^*$ moves  with a constant velocity,
so that its true trajectory passes near the optical
axis.
 We take the central mass to be $M=2.5\times 10^6\;
M_{\odot}$, and the distance to it $D_{ol}=8$ kpc.
The compact dark object is modeled as a relativistic degenerate
fermion star
with the fermion mass $m_{F}=15$ keV  and the degeneracy factor $g=4$.
We compare the results of this model  with those of the black hole
of equal mass.
From the observational point of view,
the black hole is essentially a pointlike object.
On the other hand, the fermion star is an extended object
with a nontrivial mass distribution and the radius
$R_F=18.52\:{\rm mpc}=0.4775\: {\rm arcsec}$.

\begin{figure}[p]
\vbox{\vskip230pt
\includegraphics{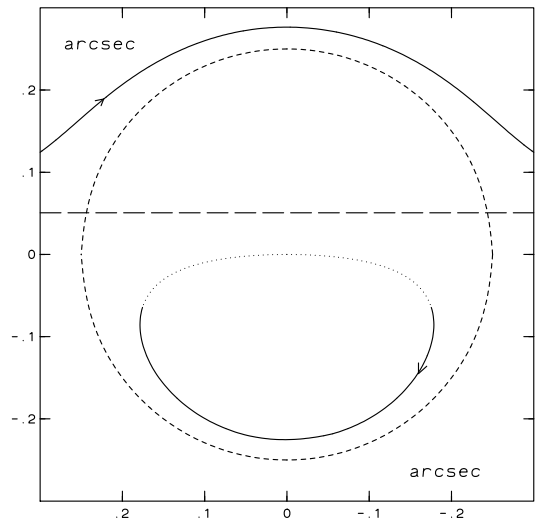}}
\caption{Trajectories of the primary and the secondary image
(solid line)
of a star
lensed by Sgr A$^*$ in the black-hole scenario.
The star is moving along a trajectory
200 pc behind Sgr A$^*$ with the impact $L=2$ mpc.
The distance to Sgr A$^*$ is assumed to be 8 kpc.
The ($x,y$) projection of the star trajectory is shown
by the horizontal dashed line.
The Einstein ring is represented by the dashed circle.
The dotted line represents the continuation of the
secondary image trajectory corresponding to the
part of the trajectory of the primary image that
goes outside the boundaries of the plot.}
\label{fig4}
\end{figure}

\begin{figure}[p]
\vbox{\vskip230pt
\includegraphics{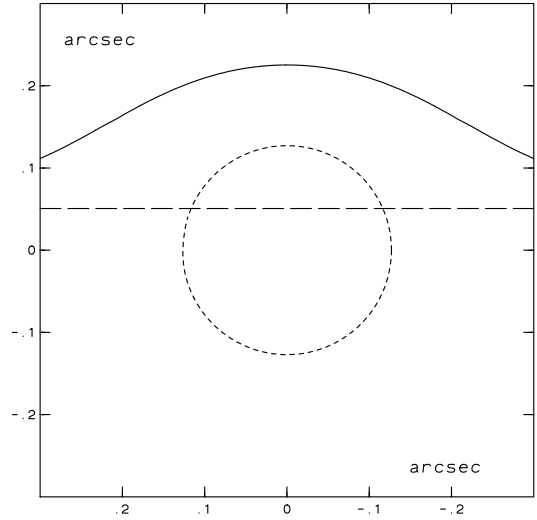}}
\caption{
Trajectory of the image
(solid line)
of a star
lensed by Sgr A$^*$ in the fermion-star scenario.
The star is moving along a trajectory
200 pc behind Sgr A$^*$ with the impact $L=2$ mpc.
The ($x,y$) projection of the star trajectory is shown
by the horizontal dashed line.
The Einstein ring is represented by the dashed circle.}
\label{fig5}
\end{figure}

\begin{figure}[p]
\vbox{\vskip230pt
\includegraphics{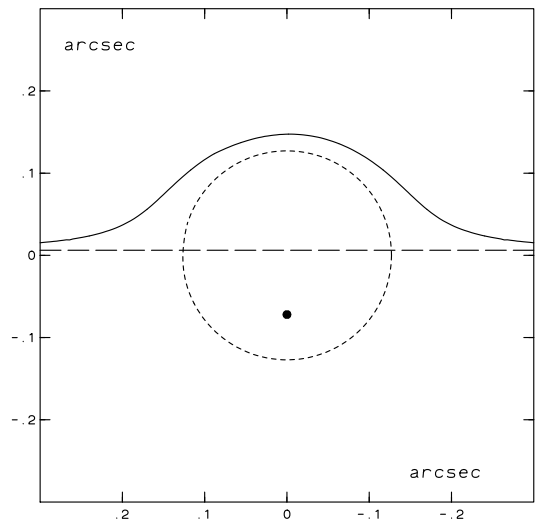}}
\caption{
Same as in Figure \protect\ref{fig5},
with $L=0.2497$ mpc.
The dot inside the Einstein ring represents
a degenerated trajectory loop of the two
secondary images.}
\label{fig6}
\end{figure}

\begin{figure}[p]
\vbox{\vskip230pt
\includegraphics{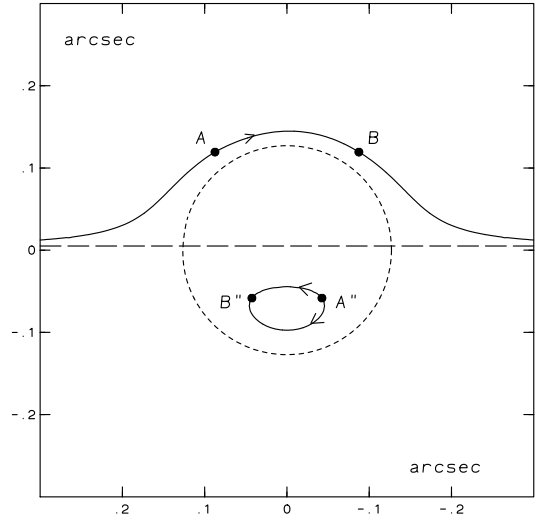}}
\caption{
Same as in Figure \protect\ref{fig5},
with $L=0.2$ mpc.
The secondary images appear at the point $A''$
and disappear at $B''$.}
\label{fig7}
\end{figure}

Let us now assume that a star at a distance $D_{ls}$ behind
 Sgr A$^*$ moves towards the optical axis
with the impact parameter $L$.
We can orient the coordinate system so that the
projected velocity $v_0$ is parallel to the $x$ axis.
In Figures
\ref{fig4},
\ref{fig5},
\ref{fig6},
and \ref{fig7}
we plot the trajectory of the images for
$D_{ls}=200$ pc and for various $L$ ranging
from 0.2 to 2 mpc.
The dashed circle
centered at Sgr A$^*$ represents
the Einstein ring,
the horizontal dashed line describes the true
trajectory of the lensed star,
and the solid line represents
the image trajectories.
The image outside the Einstein ring is referred to as
{\em primary} and those inside as {\em secondary}.
Figure \ref{fig4} shows the trajectories of images
lensed by the black hole for $L=2$ mpc.
The solid line outside the Einstein ring
represents the trajectory of the primary image
which begins to deviate substantially from
the true trajectory as it approaches the Einstein ring.
The trajectory of the secondary image begins at
the  point ($x= -3 R_{\rm S}/2, y=0$)
at the time when the primary image is at infinity on the left.
As the star approaches the optical axis and passes to the right,
the secondary image makes a loop and finishes at
the  point ($x= 3R_{\rm S}/2, y=0$)
 symmetrical to the initial point.
In addition to the secondary image, the black hole lens possesses
an infinite series of images, very close to the circle of
radius
$ 3 R_{\rm S}/2$,
that are much fainter and practically
unobservable.

In contrast to the black hole, which always produces multiple
images,
a transparent extended object, like a fermion star, may have
one, two, or three images depending on $L$.
For $L=2$ mpc, only one image exists as shown in Figure
\ref{fig5}, and its trajectory is similar to the
primary image trajectory in the case of black hole.
For small values of $L$ , less than 0.2497 mpc,
two secondary images will appear at the point $A''$
corresponding to the position $A$ of the primary
image (Figure \ref{fig7}).
As the primary image approaches the point
$B$, the two secondary images coalesce and disappear
at the point $B''$.
The value $L=0.2497$ is special as, in that case,
the loop made by the two secondary images
shown in Figure \ref{fig7} will degenerate to
a very bright single image that will appear
(and disappear) when the primary image
reaches the top of its trajectory (Figure \ref{fig6}).

\begin{figure}[p]
\vbox{\vskip230pt
\includegraphics{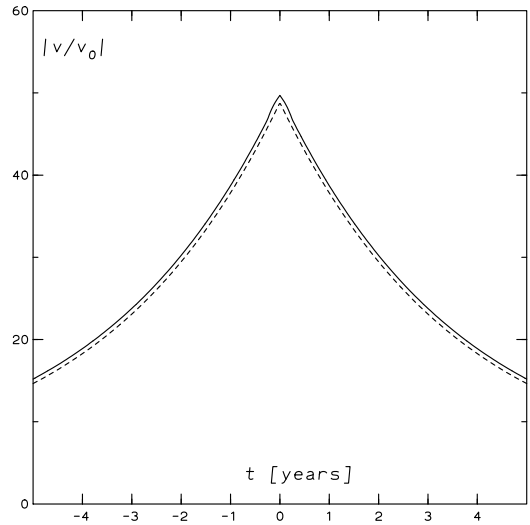}}
\caption{
Velocity of the primary (solid line) and the
secondary (dashed line) image divided by
the velocity of the star,
for lensing by a black hole as in
Figure \protect\ref{fig4} with $L=0.2$ mpc.}
\label{fig8}
\end{figure}

\begin{figure}[p]
\vbox{\vskip230pt
\includegraphics{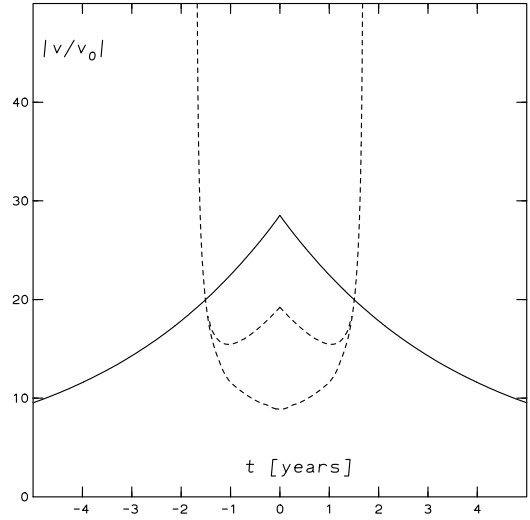}}
\caption{
Velocity of the primary (solid line) and the
secondary (dashed line) images divided by
the velocity of the star,
for lensing by a fermion star as in
Figure \protect\ref{fig7}.}
\label{fig9}
\end{figure}

One  important observable that may significantly differ
between
the two models of Sgr~A$^*$
is the velocity of the images.
Using  simple geometric
considerations, we find
the square of the image velocity $v$ divided  by
the projected velocity of the star $v_0$ in terms of the radial and
tangential magnifications
\begin{equation}
(v/v_0)^2=\mu_r^2 (1-z^2)+
\mu_t^2 z^2,
\end{equation}
where $z$ is the ratio of the impact parameter
to the
angular position of the star,
 $z=L/(\beta D_{os})$.
In Figures \ref{fig8} and \ref{fig9} we plot the absolute value $|v/v_0|$
as a function of time,
assuming that the star moves with the velocity
$v_0=50$ km/s $= 1.286$ mas/year, typical of the velocity of a star in
the Galaxy.
The  behavior of the primary image is similar
in the two models, whereas the behavior of the secondary image
is different.
The secondary image lensed by  the black hole
(Figure \ref{fig8})
moves in the opposite direction with almost exactly
the same velocity as the primary one.
For the fermion star
(Figure \ref{fig9}),
the time dependence of the velocities of
the two secondary images
is quite different
compared with the velocity of the primary image.
Note that during the period of about four years,
when all the three images are present,
both the primary image and the secondary images
are by an order of magnitude faster than the
lensed star itself.

\begin{figure}[p]
\vbox{\vskip230pt
\includegraphics{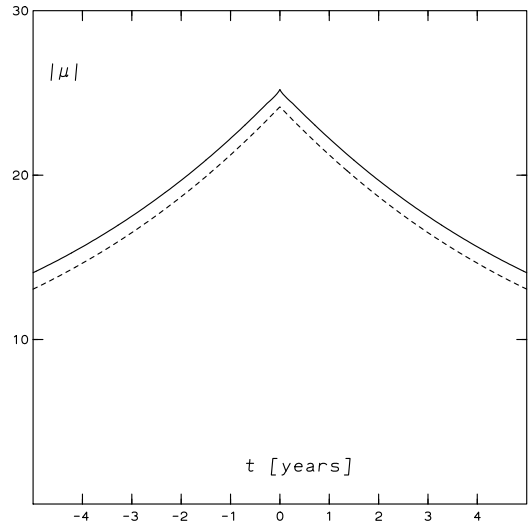}}
\caption{
Magnification of the primary (solid line) and the
secondary (dashed line) image
as a function of time
for lensing by a black hole as in
Figure \protect\ref{fig4} with $L=0.2$ mpc.}
\label{fig10}
\end{figure}

\begin{figure}[p]
\vbox{\vskip230pt
\includegraphics{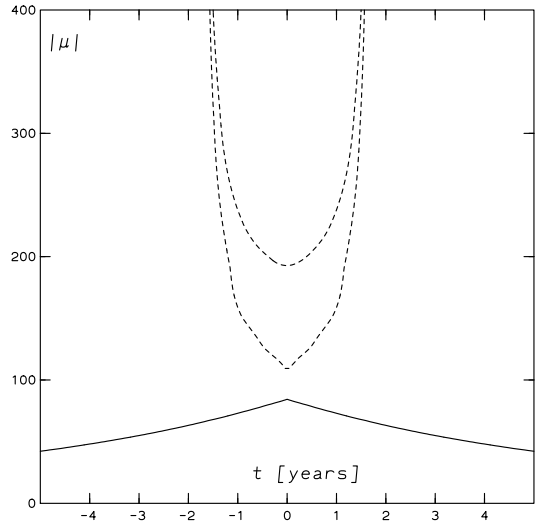}}
\caption{
Magnification of the primary (solid line) and the
secondary (dashed line) images
as a function of time
for lensing by a fermion star as in
Figure \protect\ref{fig7}.}
\label{fig11}
\end{figure}

In Figures \ref{fig10} and \ref{fig11}  we plot
 the total magnification  of
the images
as a function of time.
Comparing these  figures, one notes  two
features that are very distinct in the two models.
 First,  for the given set of parameters,
the magnifications of the two images lensed by the black hole
are almost the same,
whereas the magnifications of the
 images lensed by the fermion star
 are very different.
Second, in the black hole case, the secondary image
is slightly fainter than the primary one
(Figure \ref{fig10}), as opposed to
fermion-star lensing where the secondary images are
always brighter
 (Figure \ref{fig11}).
The magnification of the secondary images becomes singular
at the points of ``creation" $A''$ and ``annihilation" $B''$.

Next, we investigate the possibility that one of the
stars in the vicinity of Sgr A$^*$,
e.g., S1 (\cite{eck96,eck97}) is a lensed image
 of an object whose
true position is behind
 Sgr A$^*$.                                                            l
 The image of S1
is seen at $\theta_{1} =\sqrt{0.19^2
+0.04^2}=0.1942\; {\rm arcsec}=9.413\times 10^{-7}\; {\rm rad}$
 (\cite{mun98}).
Since the closest distance between the primary
 ray of light and the center
of the deflecting object is
$r_{0(1)} =D_{ol} \theta_{1}$,
we can calculate
 from (\ref{alkap})
the deflection angle of the primary ray.
 For the black hole,
we find $\hat{\alpha}_{1}=6.347\times 10^{-5}\; {\rm rad}$ and
for the fermion star,
 $\hat{\alpha}_{1}=3.264\times 10^{-5}\; {\rm rad}$.
In both cases, $\hat{\alpha}_{1}>\theta_{1}$,
implying
that in both  the
 black-hole and fermion-star   scenarios
the extended ray crosses the optical axis
behind the lens.
The crossing occurs at a distance
\begin{equation}\label{D0}
 D_0 =r_{0(1)}/\tilde{\theta} \;
\end{equation}
from the center of the lens, where $\tilde{\theta}=
\hat{\alpha}_{1}-\theta_{1}$.
The true position of the source is
somewhere on the extended ray.

\begin{figure}[p]
\vbox{\vskip230pt
\includegraphics{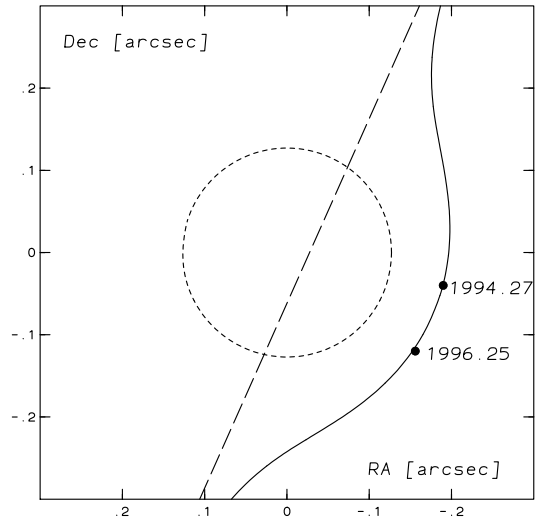}}
\caption{
Trajectory of the image
(solid line)
of S1
lensed by Sgr A$^*$ assuming the fermion-star scenario.
The star is moving along a trajectory (dashed line)
205 pc behind Sgr A$^*$ with the impact $L=1$ mpc.
The Einstein ring is represented by the dashed circle.}
\label{fig12}
\end{figure}

\begin{figure}[p]
\vbox{\vskip230pt
\includegraphics{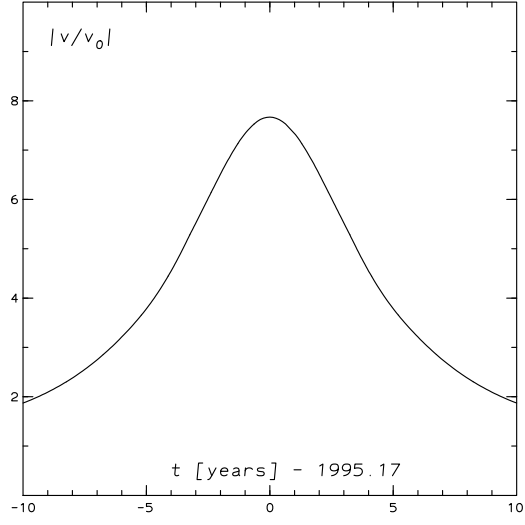}}
\caption{
Ratio of the
image velocity of S1
to
the velocity of the star as a function of time.}
\label{fig13}
\end{figure}

In Figures \ref{fig12} and
\ref{fig13} we present hypothetical lensing of
S1.
In Figure \ref{fig12} we fit the  two positions
observed in 1994 and 1996 (\cite{eck97}), with a trajectory
of the primary image that is close to the
Einstein ring, and assuming that no secondary
image appears (a situation represented by
Figure \ref{fig5}).
With only two points, such a fit is, of course, not unique.
Our fit is made with the
impact parameter $L=1$ mpc and the distance $D_{ls}=205$ pc.
We have made this choice of $L$ and
$D_{ls}$ in order to simply demonstrate that
a high velocity of one or more objects observed in the vicinity
of Sgr A$^*$ may be due to lensing.
In Figure \ref{fig13} we show
the velocity plot
corresponding to the trajectory in Figure \ref{fig12}.
The origin is placed at the middle point
 between 1994.27 and 1996.25.
 During this period the average projected image velocity
 of $1660\pm 400$ km/s
 has been deduced from the data (\cite{eck97}).
 The true velocity of the star $v_0$
is smaller by about a factor of eight,
 hence $v_0=200\pm 50$ km/s.

We have to point out that it is very unlikely
that the star S1 is indeed a lensed image.
The likelihood of such a lensing event
may be estimated
using the empirical density of stars
near the Galactic center (\cite{gen87}):
\begin{equation}\label{dens}
 \rho=2.3 \times 10^5 r^{-1.85}\;M_{\odot}/{\rm pc}^3,
\end{equation}
where $r$ is the distance from the center in pc.
By integrating this density in the tube of
radius $ L=1$ mpc that stretches from $r_1=100$ to $r_2=300$ pc
we find the number of solar masses in that region
to be of the order 0.01.
Using kinetic theory, the number of stars crossing per unit time
into that region can be estimated as
\begin{equation}\label{rate}
 \frac{dN}{dt}=\frac{1}{6}\bar{n} \bar{v}\, 2\pi L (r_2-r_1)
\end{equation}
where $\bar{n}\approx 10$ pc$^{-3}$ is the average
star density and $v_0\approx 50$ km/s the
average velocity. This gives a lensing rate of
$dN/dt \approx 10^{-4}$ yr$^{-1}$.
Hence, the probability of observing a lensing event
similar to that described above is quite small.
\section{Conclusions}
\label{conclude}
We have investigated possible gravitational
lensing effects caused by a supermassive dark
object
made of weakly interacting fermions.
Such fermion stars may be alternative candidates for
compact dark objects at galactic centers,
in particular, at the center of our galaxy
referred to as Sgr A$^*$.
We have compared  lensing effects by Sgr A$^*$
in the black hole and fermion star scenarios.
We have found a clear distinction between lensing
patterns in the two scenarios, assuming the existence of a lensed
object that moves close to the optical axis at a distance
 of about 200 pc behind Sgr A$^*$.
 We have also investigated the possibility that one of the
 rapidly moving stars near the Galactic center
 is a lensed image of an object whose true position is
 behind Sgr A$^*$.

\acknowledgments

 This work  was supported by
 the Foundation for Fundamental Research (FFR) and
 the Ministry of Science and Technology of the
 Republic of Croatia under Contract
 No. 00980102.

\clearpage


\end{document}